\documentclass[11pt]{article}
\usepackage{amsbsy,amsthm,amsmath,amssymb}
\usepackage{algorithm,algpseudocode}
\usepackage{bookmark}
\usepackage{booktabs}
\usepackage{graphicx}
\usepackage{hyperref}

\graphicspath{{figures/}}

\def\sgn{\mathop{\rm sgn}\nolimits}

\def\dist{\mathop{\rm dist}\nolimits}

\def\argmin{\mathop{\rm argmin}\nolimits}

\newcommand{\bzero}{\boldsymbol{0}}

\newcommand{\bp}{\boldsymbol{p}}

\newcommand{\bs}{\boldsymbol{s}}

\newcommand{\bu}{\boldsymbol{u}}
\newcommand{\bv}{\boldsymbol{v}}

\newcommand{\bx}{\boldsymbol{x}}
\newcommand{\by}{\boldsymbol{y}}
\newcommand{\bz}{\boldsymbol{z}}
\newcommand{\bA}{\boldsymbol{A}}

\newcommand{\bC}{\boldsymbol{C}}
\newcommand{\bD}{\boldsymbol{D}}

\newcommand{\bI}{\boldsymbol{I}}

\newcommand{\bK}{\boldsymbol{K}}

\newcommand{\bU}{\boldsymbol{U}}
\newcommand{\bV}{\boldsymbol{V}}

\newcommand{\bX}{\boldsymbol{X}}
\newcommand{\bY}{\boldsymbol{Y}}

\newcommand{\balpha}{\boldsymbol{\alpha}}
\newcommand{\bbeta}{\boldsymbol{\beta}}

\newcommand{\bSigma}{\boldsymbol{\Sigma}}

\newcommand{\Real}{\mathbb{R}}

\title{Algorithms for Sparse Support Vector Machines}
\author{Alfonso Landeros$^{1}$, Kenneth Lange$^{1,2,3}$
\\
\\
\\
Departments of Computational Medicine$^{1}$, \\
Human Genetics$^{2}$, and Statistics$^{3}$ \\
University of California \\
Los Angeles, CA 90095\\
\\
\\
\\
\\
Research supported in part by
\\ USPHS grants R35 GM141798 and HG006139.}

\begin{document}

\maketitle
\newpage 

\section*{Abstract}

Many problems in classification involve huge numbers of irrelevant features. Model selection reveals the crucial features, reduces the dimensionality of feature space, and improves model interpretation. In the support vector machine literature, model selection is achieved by $\ell_1$ penalties. These convex relaxations seriously bias parameter estimates toward 0 and tend to admit too many irrelevant features. The current paper presents an alternative that replaces penalties by sparse-set constraints. Penalties still appear, but serve a different purpose. The proximal distance principle takes a loss function $L(\bbeta)$ and adds the penalty $\frac{\rho}{2}\dist(\bbeta, S_k)^2$ capturing the squared Euclidean distance of the parameter vector $\bbeta$ to the sparsity set $S_k$ where at most $k$ components of $\bbeta$ are nonzero. If $\bbeta_\rho$ represents the minimum of the objective $f_\rho(\bbeta)=L(\bbeta)+\frac{\rho}{2}\dist(\bbeta, S_k)^2$, then $\bbeta_\rho$ tends to the constrained minimum of $L(\bbeta)$ over $S_k$ as $\rho$ tends to $\infty$. We derive two closely related algorithms to carry out this strategy. Our simulated and real examples vividly demonstrate how the algorithms achieve much better sparsity without loss of classification power.

\bigskip

\noindent
\textbf{keywords:} sparsity; discriminant analysis; unsupervised learning; Julia

\newpage

\section{Introduction}

Support vector machines (SVMs) are powerful pattern recognition tools \cite{cortes1995supportvector} with a wide range of applications across machine learning and statistics. Success stories in supervised learning include optical character recognition \cite{decoste2002training}, image segmentation \cite{barghout2015spatialtaxon}, text categorization \cite{joachims1998text,pradhan2004shallow}, protein structure prediction \cite{dunbrack2006sequence}, and early detection and classification of human cancers \cite{sewak2007svm}. It is possible to extend SVM to online algorithms for resource-limited computing environments \cite{cauwenberghs2000incremental,laskov2006incremental} and beyond classification to unsupervised learning \cite{ben-hur2002support}.

Many variations on supervised SVMs have been proposed. For example, the classic soft-margin classifier features a $\ell_{2}$ penalty term on model coefficients, $\sum_{j} \beta_{j}^{2}$, that regularizes classification. To induce model selection, this ridge penalty has since been amended to an $\ell_{1}$ penalty \cite{zhu20031norm}. Other loss functions, such as quadratic and Huber hinge errors, have been proposed as alternatives to the typical hinge loss $u_+=\max\{0,u\}$ to promote better prediction and robustness to outliers \cite{groenen2008svmmaj}.

The aim of this work is to present a flexible framework that attacks sparsity directly and generates attractive search algorithms. We focus on primal problems under the squared-hinge loss with extensions to kernel machines for nonlinear data. Our major contribution is to replace sparsity inducing penalties with sparsity constraint sets. This goal is achieved by estimating parameters by an iterative proximal distance algorithm \cite{lange2016mma}. Our proximal distance algorithm comes in two closely related flavors. Both exhibit comparable, and often superior, predictive accuracy to existing SVM approaches. Although our proximal distance algorithms are sometimes slower than competing methods, the output of both is easier to interpret and better at revealing the sparse signals hidden in high-dimensional data.

\section{Motivation}

In binary classification we adopt the squared hinge-loss $\max\{0,1-u\}^{2}$ in the criterion 
\begin{equation}
    \label{eq:basic-primal-model}
    L(\bbeta \mid \by, \bX) = \frac{1}{2n} \sum_{i=1}^{n} \max\{0, 1-y_{i} \bx_{i}^{\top} \bbeta \}^{2}.
\end{equation}
Here the $n$ observations $(y_{i}, \bx_{i})$ consist of a class label $y_{i} \in \{-1,1\}$ and a feature vector $\bx_{i} \in \Real^{p+1}$.
The parameter vector $\bbeta \in \Real^{p+1}$
defines the hyperplane separating the two classes. The last component $\beta_{p+1}$ of $\bbeta$ represents the model intercept; the last component $x_{i, p+1}$ of each $\bx_i$ is accordingly 1. The form of the loss allows for a small margin of error in classification.  

Rather than directly minimize $L(\bbeta \mid \by, \bX)$ defined by equation (\ref{eq:basic-primal-model}), we turn to the MM principle \cite{lange2016mma,lange2000optimization} and invoke the quadratic majorization
\[
    \max\{0, 1-u\}^{2} \le \begin{cases}
        (u_{m} - u)^{2}, & u_{m} \ge 1 \\
        (1 - u)^{2}, & u_{m} < 1 
    \end{cases}
\]
at iteration $m$ suggested by Groenen et al \cite{groenen2008svmmaj}. Note the two sides of the majorization agree when $u=u_m$. Given that all $y_{i}^{2} = 1$, the term by term application of the majorization creates the overall quadratic surrogate
\begin{equation}
    \label{eq:basic-surrogate}
    g(\bbeta \mid \bbeta_{m})
    = \frac{1}{2n} \| \bz_{m} - \bX \bbeta \|^{2},
    \qquad
    \bz_{mi} = \begin{cases}
        \bx_{i}^{\top} \bbeta_{m} & y_{i} \bx_{i}^{\top} \bbeta_{m} \ge 1 \\
        y_{i} & y_{i} \bx_{i}^{\top} \bbeta_{m} < 1
    \end{cases}.
\end{equation}
This maneuver reduces the original minimization problem to sequence of easier minimization problems that can be solved by iteratively-reweighted least squares (IRLS); specifically, $\bbeta_{m+1} = \argmin_{\bbeta}~\frac{1}{2n} \| \bz_{m} - \bX \bbeta \|^{2}$. The MM principle implies that
every iteration decreases the objective function (\ref{objective_function}). Nguyen and McLachlan also apply the MM principle to support vector machines and make the connection to IRLS, albeit with a different majorization \cite{nguyen2017iterativelyreweighted}. The surrogate (\ref{eq:basic-surrogate}) is nicer because all weights in IRLS remain 1. 

\section{Sparse SVMs via Distance Penalization}

The convex surrogate (\ref{eq:basic-surrogate}) does not admit a unique solution in general unless its Hessian $\bX^{\top} \bX$ is positive definite. The standard approach in the literature augments the (squared) hinge-loss  (\ref{eq:basic-primal-model}) with the ridge penalty $\lambda \sum_{j=1}^{p} \beta_{j}^{2}$, omitting the intercept. Putting $\lambda > 0$ (a) induces an upper bound of $\lambda^{-1}$ on the margin separating distinct classes in the original problem, (b) enforces strict convexity, and (c) guarantees a unique minimum of both the objective and surrogate. Other regularizations are tractable and achieve the same goal \cite{groenen2008svmmaj}.

Here we focus on inducing feature selection in soft-margin classifiers. In brief, we impose a set constraint on the loss (\ref{eq:basic-primal-model}) that directly controls sparsity in $\bbeta$. The set $S_{k} \subset \Real^{p+1}$ determined by the hyperparameter $k$ limits $\bbeta$ to have at most at most $k$ nonzero values among its first $p$ components. Fortunately, projection onto $S_{k}$ is straightforward, and the hyperparameter choices $k \in \{0,1,2,\ldots,p\}$ are finite in number.  The projection operator $P_{S_k}(\bbeta)$ sets to zero all but the largest $k$ entries in magnitude of $\beta_1$ through $\beta_p$. This goal can be achieved efficiently by a partial sort of these entries. The intercept is ignored  by projection, and the value of $k$ may be chosen by cross-validation.

Following the penalty method of constrained optimization \cite{beltrami1970algorithmica,courant1943variationala}, we minimize the continuously differentiable but unconstrained objective
\begin{equation}
    f_\rho(\bbeta) = L(\bbeta \mid \by, \bX) + \frac{\rho}{2(p-k+1)} \dist(\bbeta, S_{k})^{2} \label{objective_function}
\end{equation}
for a large value of the annealing parameter $\rho \ge 0$. The distance penalty enforces near sparsity. In the limit as $\rho$ tends to $\infty$, the solution vector $\bbeta_\rho$ tends to the solution of the constrained problem. We attach the prefactor $(p-k+1)^{-1}$ to account for the number of nonzero entries in the difference $\bbeta - P_{S_{k}}(\bbeta)$. This choice facilitates comparisons between different sparsity levels $k$. Combining our previous majorization (\ref{eq:basic-surrogate}) with distance majorization
\[
    \dist(\bu, S) \le \|\bu - P_{S}(\bu_{m})\|^{2},
\]
yields the sum of squares surrogate
\begin{equation}
    \label{eq:generic-surrogate}
    g_{\rho}(\bbeta \mid \bbeta_{m})
    =
    \frac{1}{2} 
    \left\|
    \begin{bmatrix}
        a \bz_{m} \\
        b \bp_{m}   
    \end{bmatrix}
    -
    \begin{bmatrix}
        a \bX \\
        b \bD
    \end{bmatrix} \bbeta
    \right\|_{2}^{2}
\end{equation}
with the choices 
\[
    a = \frac{1}{\sqrt{n}}, \quad
    b = \sqrt{ \frac{\rho}{p-k+1} }, \quad
    \bp_{m} = P_{S_{k}}(\bbeta_{m}), \quad
    \bD = \bI_{p \times p}.
\]
In this form it is clear that our model formulation contains the $\ell_{2}$ penalty as a special case; simply substitute $\bp_{m} = \bzero$ and $b = \sqrt{\lambda}$. Moreover, the MM principle restores differentiability and permits the use of exotic regularization penalties with computable projection operators. Although convexity is lost, sparsity constraints permit identification of features driving a classifier's decision boundary. Our previous experience \cite{keys2019proximala, landeros2020extensions,xu2017generalizeda} supports the value of the proximal distance principle in building parsimonious models with non-convex set constraints.

While soft-margin classifiers are known to perform decently even on data that are not linearly separable, they do not generalize well to inherently nonlinear data. Instead, one can invoke the``kernel trick'' via the representation $\bbeta = \sum_{i} y_{i} \alpha_{i} \phi(\bx_{i})$ with possibly infinite dimensional feature map $\phi(\bx)$. Substituting this representation into equation (\ref{eq:basic-primal-model}) yields the loss
\[
    L(\balpha \mid \by, \bX) = \frac{1}{2n} \sum_{i=1}^{n} \max\Big\{0, 1
        -y_{i} \sum_{j=1}^{n} y_{j} \alpha_{j} K_{ij}\Big\}^{2}.
\]
Applying our previous majorization yields the same surrogate (\ref{eq:basic-surrogate}) but with $\bX$ replaced by the matrix $\bK \bY$ and $\bbeta$ replaced by the vector $\balpha \in \Real^{n}$. Here $K_{ij} = \langle \phi(\bx_{i}), \phi(\bx_{j}) \rangle$ defines a positive definite kernel and $\bY = \mathrm{Diag}(\by)$. Unfortunately, feature selection is generally impossible under this representation. Sparsity constraints instead become a mechanism for controlling the number of support vectors (representative samples). We focus our attention on the $L(\bbeta)$ model rather than $L(\balpha)$ model, with the understanding that one can readily pass to the kernel version as needed.

\section{Algorithms}

The gradient of the surrogate (\ref{eq:basic-surrogate}) reads
\begin{equation}
    \label{eq:surrogate-gradient}
    \nabla g_{\rho}(\bbeta \mid \bbeta_{m}) = 
    (a^{2} \bX^{\top} \bX + b^{2} \bI) \bbeta 
    - a^{2} \bX^{\top} \bz_{m} - b^{2} \bp_{m}.
\end{equation}
Taking the thin SVD $\bX = \bU \bSigma \bV^{\top}$ of $\bX$ enables one to directly solve the stationary equation $\nabla g_{\rho}(\bbeta \mid \bbeta_{m}) = \bzero$.
In the appendix we derive the iteration map
\begin{eqnarray*}
    \bbeta_{m+1}
    &=& \bp_{m} + a^{2} \bV (a^{2} \bSigma^{2} + b^{2} \bI)^{-1} \left( 
        \bSigma \bU^{\top} \bz_{m} - \bSigma^{2} \bV^{\top} \bp_{m}
     \right) \\
    &=& \bp_{m} + \sum_{j=1}^{r}
        \frac{ a^{2} }{ a^{2} s_{j}^{2} + b^{2} } \left[ 
            s_{j} \bu_{j}^{\top} \bz_{m} - s_{j}^{2} \bv_{j}^{\top} \bp_{m}
        \right] \bv_{j}.
\end{eqnarray*}
The second line of this MM update can be implemented as described in Algorithm \ref{alg:MM}.  Each update step requires one copy to initialize $\bbeta_{m+1}$ as $\bp_m$ followed by two dot products (sizes $n$ and $p+1$) and one BLAS \texttt{AXPY} per singular value. The time complexity of each iteration therefore scales as $\mathcal{O}[2nr + (2r+1)(p+1)]$. Alternatively, one may implement the updates based on the first line above using BLAS Level 2 operations at the cost of extra memory allocations.

Let us highlight a few of the merits of the MM algorithm. The Gram matrix $\bX^\top\bX$ is never formed. The row and column dimensions of $\bX$ are decoupled so one can expect favorable performance for many high-dimensional problems. Fortunately, the expensive SVD of $\bX$ need only be extracted once and applies across all values of $\rho$ and all sparsity levels $k$. This allows one to solve each IRLS subproblem quickly. As already noted, the MM algorithm leads to a steady decrease in the objective $f_\rho(\bbeta)$ for $\rho$ fixed. The shrewd reader may observe that the initial value $\bbeta_{0}$, the choice of annealing schedule sending $\rho \to \infty$, and the non-uniqueness of the projection operator may impact the the quality of the ultimate solution. The second of these three concerns is the most worrisome. Non-uniqueness almost never occurs, and gentle annealing usually weaves its way to a good solution. On the other hand, too gentle annealing wastes time.

\begin{algorithm}[htp]
\small
\caption{MM}
\label{alg:MM}
\begin{algorithmic}[1]
    \Require $\bU, \bV, \bs$ such that $\bX = \bU \mathrm{Diag}(\bs) \bV^{\top}$
    \State Set $a^{2} \gets 1/n$ and $b^{2} \gets \rho / (p-k+1)$.
    \While {not converged}
    \State   Set $z_{mi} \gets 
    \bx_{i}^{\top} \bbeta_{m} \: \text{if} \: y_{i} \bx_{i}^{\top} \bbeta_{m} \ge 1    \: \text{and} \: y_i \: \text{otherwise}$
    for $i = 1,2,\ldots,n$.
    \State Project $\bp_{m} \gets P_{S_{k}}( \bbeta_{m} )$.
    \State Copy $\bbeta_{m+1} \gets \bp_{m}$.
    \For {$j = 1,2,\ldots,r$}
        \State Evaluate $uz = \bu_{j}^{\top} \bz_{m}$ and $vp = \bv_{j}^{\top} \bp_{m}$.
        \State Compute $c_{1} = \frac{ a^{2} s_{j} }{ a^{2} s_{j}^{2} + b^{2} }$,
            $c_{2} = \frac{ a^{2} s_j^{2} }{ a^{2} s_{j}^{2} + b^{2} }$, and
            $c_{3} = c_{1} uz - c_{2} vp$.
        \State Accumulate $\bbeta_{m+1} \gets \bbeta_{m+1} + c_{3} \bv_{j}$.
    \EndFor
    \EndWhile
\end{algorithmic}
\end{algorithm}

Alternatively, the tangency condition $\nabla  g_{\rho}(\bbeta_{m} \mid \bbeta_{m}) = \nabla f_\rho(\bbeta_m)$ suggests implementing gradient descent. Minimizing the surrogate $g_{\rho}(\bbeta \mid\bbeta_{m})$ in the direction $-\nabla f_\rho(\bbeta_m)$ leads to the update
\begin{eqnarray*}
\bbeta_{m+1} & = & \bbeta_{m} - t_{m} 
\nabla f(\bbeta_m) \\
t_{m} & = & \frac{\|\nabla f_\rho(\bbeta_m)\|^{2} }{
        a^{2} \|\bX \nabla f_\rho(\bbeta_m)\|^{2} + b^{2} \|\nabla f_\rho(\bbeta_m)\|^2}
\end{eqnarray*}
derived in the appendix. The fact that the surrogate is quadratic and strictly convex in $\bbeta$ implies that $\bbeta_{m+1}$ furnishes its exact minimum along the steepest descent direction. Once again, minimizing the surrogate forces a decrease in $f_\rho(\bbeta)$. In practice, it is helpful to add a small perturbation to the denominator of $t_{m}$ to guard against the indeterminate form $0 / 0$. The convergence properties of both the MM and SD algorithms are shared by more general proximal distance algorithms and sketched in the references \cite{keys2019proximala,lange2021nonconvex}.
Algorithm \ref{alg:SD} summarizes the SD algorithm in pseudocode. 

Our steepest descent algorithm has several advantages. First, the updates of Algorithm \ref{alg:SD} are simpler than those of its sibling Algorithm \ref{alg:MM}. Second, changing the hyperparameters $\rho$ and $k$ imposes no additional overhead. Third, gradient descent makes little memory demand beyond storing input data and model coefficients. Fourth, the quadratic surrogate relies on an exact line search that implicitly incorporates curvature information from the objective. Fifth, the method is simple to implement and can run on both CPUs and GPUs.

\begin{algorithm}[htp]
\small
\caption{SD}
\label{alg:SD}
\begin{algorithmic}[1]
    \State Set $a^{2} \gets 1/n$ and $b^{2} \gets \rho / (p-k+1)$.
    \While {not converged}
        \State Set  $v_{i} = - a^{2} y_{i} \max\{0, 1-y_{i} \bx_{i}^{\top} \bbeta_{m}\}$ for $i = 1,2,\ldots,n$.
        \State Project $\bp_{m} \gets P_{S_{k}}( \bbeta_{m} )$.
        \State Evaluate $\nabla f_\rho(\bbeta_{m}) \gets b^{2} [ \bbeta_{m} - \bp_{m} ]
            + \bX^{\top} \bv$.
        \State Compute $t_{m}$ as described.
        \State Iterate $\bbeta_{m+1} \gets \bbeta_{m} - t_m \nabla f_\rho(\bbeta_{m})$.
    \EndWhile
\end{algorithmic}
\end{algorithm}

We conclude this section by pointing out that both proximal distance algorithms benefit from  Nesterov acceleration  \cite{nesterov1983method}.
Starting with $j=1$, each update $\bbeta_{m+1}$ can be potentially be replaced according to the rule
\[
    \bbeta_{m+1} \mapsto
    \bbeta_{m+1} + \frac{j-1}{j+r-1} [\bbeta_{m+1} - \bbeta_{m}],
\]
where $r=3$ is the preferred choice \cite{su2014differential}. Because Nesterov
acceleration does not preserve the descent property, we omit it in the early iterations of the MM and SD algorithms.  This actions lets the algorithms hit a neighborhood of the optimal point before we aggressively accelerate.

\section{Numerical Experiments}

Here we apply our algorithms to supervised classification involving our own synthetic examples and datasets from the UCI Machine Learning Repository \cite{dua2019uci}.
The chosen datasets, which address both overdetermined and underdetermined problems, are listed in Table \ref{tab:1}.
\begin{table}[htp]
    \centering
    \scriptsize
    \begin{tabular}{llllll}
    \toprule
    Name & Classes & Samples & Features & No. Train & No. Test \\
    \midrule
    \texttt{synthetic} & 2 & 1000 & 500 & 800 & 200 \\
    \texttt{iris} & 3 & 150 & 4 & 120 & 30 \\
    \texttt{letter-recognition}$\ast$ & 26 & 20000 & 16 & 16000 & 4000 \\
    \texttt{splice}$\ast$ & 3 & 3186 & 180 & 2549 & 637 \\
    \texttt{spiral} & 3 & 1000 & 2 & 800 & 200 \\
    \texttt{optdigits}$\dagger$ & 10 & 5620 & 64 & 3822 & 1798 \\
    \texttt{breast-cancer-wisconsin}$\dagger$ & 2 & 683 & 9 & 546 & 137 \\
    \texttt{TCGA-PANCAN-HiSeq}$\ast$ \cite{weinstein2013cancer} & 5 & 801 & 10000 & 641 & 160 \\
    \bottomrule
\end{tabular}
    \caption{
        \label{tab:1}
        Summary of datasets for numerical experiments.
        Data are standardized to have mean 0 and variance 1 unless otherwise stated. The symbol $\ast$ indicates data are normalized using the min-max transformation;  $\dagger$ indicates data are not transformed.
        }
\end{table}
\noindent Let us briefly elaborate on our two synthetic examples:
\begin{itemize}
    \item[(a)] \texttt{synthetic}: Targets are simulated as $y_i = \sgn{\bx_i^\top \bbeta_{0}}$ with $\bx_i$ a sample from $\mathcal{N}(\bzero, \bSigma)$. There are two causal predictors $\beta_{2} = -\beta_{1} = 10$; otherwise $\beta_{j} = 0$.
    The covariance structure of $\bx_i$ is chosen so that only the two informative features $(x_{1}, x_{2})$ are strongly (positively) correlated. All remaining pairs $(x_{i}, x_{j})$ are uncorrelated.

    \item[(b)] \texttt{spiral}: The data consist of three noisy spirals with 600, 300, and 100 samples in each class and variances 0.01, 0.04, and 0.09, respectively.
    These data are highly nonlinear, so we evaluate our algorithms using a Gaussian kernel version of SVM classification.
\end{itemize}

We use the standard decision rule $\bx \mapsto \sgn (\bx^{\top} \bbeta$) for binary classification. In examples with $c > 2$ classes, we train $\binom{c}{2}$ SVMs under the one-versus-one paradigm \cite{hsu2002comparison}. Briefly, if classes are labeled $\{1, 2, \ldots, c\}$, the decision rule is $\bx \mapsto \arg\max v_{j}$, where $v_{j}$ is the number of votes cast for class $j$ when class $j$ is compared to every other class $i$.

Algorithms \ref{alg:MM} and \ref{alg:SD} are referred to as MM and SD, respectively. Our implementations of MM and SD use Nesterov acceleration with restarts. In both cases, the algorithm is run with fixed $\rho$  until the following convergence criterion is met
\[
    \|\nabla f_{\rho}(\bbeta_{m+1})\|^{2} < \epsilon
    \quad \text{or} \quad
    m+1 > m_{\text{inner}}.
\]
where $f_\rho(\bbeta)$ is the penalized objective (\ref{objective_function}).
We fix $\epsilon = 10^{-6}$ and allow a maximum $m_{\text{inner}} =10^{4}$ or $10^{5}$ iterations across all examples. This rule ensures that an algorithm makes sufficient progress towards minimizing $f_\rho(\bbeta)$. In practice, we observe that both algorithms require far fewer than $10^5$ iterations on a given subproblem. Starting from $\rho=1$, we update $\rho \to M \times \rho$ for some multiplier $M>1$ every $i_{\text{outer}}=100$ outer iterations; thus, the maximum value of $\rho$ is $M^{100}$. Taking $d_{i} = (n-k+1)^{-1} \dist(\bbeta_{i}, S_{k})^{2}$ as the distance at outer iteration $i$, each algorithm halts before all $100$ outer whenever
\[
    d_{i} \le 10^{-6}
    \quad \text{or} \quad
    |d_{i} - d_{i-1}| < 10^{-6} (1 + d_{i-1}).
\]
This ensures solutions are close to the chosen constraint set. These heuristic stopping rules strike a balance between delivering accurate solutions and avoiding expensive iterations that make little progress in minimization. Taking all
100 outer iterations indicates a poor annealing schedule or slow convergence. The general procedure is summarized in Algorithm \ref{alg:pdi}.
\begin{algorithm}[htp]
\scriptsize
\caption{Proximal Distance Iteration}
\label{alg:pdi}
\begin{algorithmic}[1]
    \State Set tolerance $\epsilon$, multiplier $a$, and sparsity level $s = 1-k/n$.
    \State Set outer iterations $i_{\mathrm{outer}}$ and inner iterations $m_{\mathrm{inner}}$.
    \State Initialize $\rho$, $\bbeta$, and $d_{0} \gets (n-k+1)^{-1}\dist(\bbeta, S_{k})^{2}$.
    \For {$i \gets 1,i_{\mathrm{outer}}$}
        \For {$m \gets 1,m_{\mathrm{inner}}$}
            \State Solve the subproblem $\bbeta_{m+1} \gets \argmin_{\bbeta} g_{\rho}(\bbeta \mid \bbeta_{m})$

            \If {$\|\nabla f_{\rho}(\bbeta_{m+1})\|^{2} < \epsilon$}
                \State Break.
            \EndIf
        \EndFor

        \State Update $\bbeta \gets \bbeta_{m}$.
        \State Set $d_{i} \gets (n-k+1)^{-1}\dist(\bbeta, S_{k})^{2}$.
        \If {$d_{i} < 10^{-6}$ OR $|d_{i} - d_{i-1}| < 10^{-6}[1 + d_{i-1}]$}
            \State Break.
        \Else
            \State $\rho \gets a \times \rho$
        \EndIf
    \EndFor
    \State Project final estimate $\bbeta \gets P_{S_{k}}(\bbeta)$.
\end{algorithmic}
\end{algorithm}

\subsection{Implementation}

We implement our algorithms and numerical experiments in the Julia programming language \cite{Julia-2017}. Our software package calls DataDeps.jl to automatically download data sets and run any preprocessing \cite{DataDeps.jl-2018}. All of our experiments were carried out on a Manjaro Linux desktop machine equipped with an Intel i9-10900KF @ 4.9 GHz (using 8 cores) and 32 GB RAM @ 3600 MHz. Intel's BLAS implementation, MKL, was used for all linear algebra subroutines.

\subsection{Sensitivity to Initial Conditions}

\begin{figure}[htp]
    \centering
    \includegraphics[width=\textwidth]{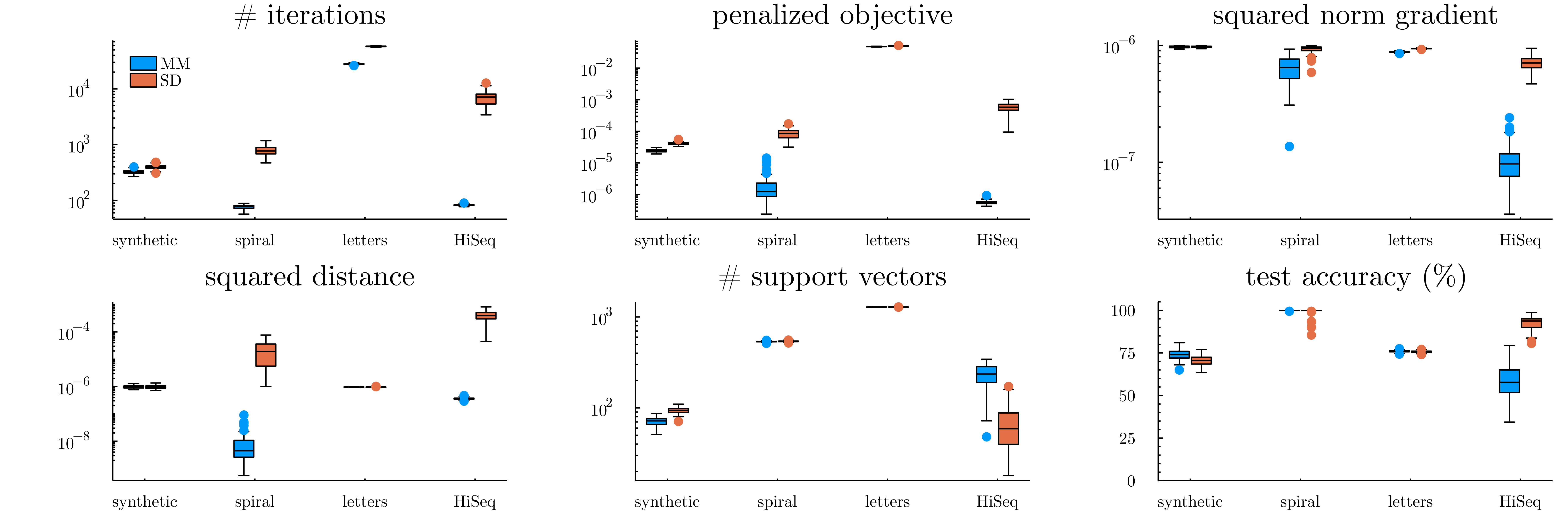}
    \caption{
        Sensitivity of algorithms to initial conditions across data sets.
        \label{fig:1}
    }
\end{figure}

Sparsity sets $S_{k}$ are closed but not convex, so imposing sparsity constraints on the original minimization problem $\min_{\bbeta} L(\bbeta)$ destroys theoretical guarantees afforded by convexity. Thus, solutions generated by both MM and SD are potentially sensitive to the initial guess $\bbeta_{0}$. To address this concern, we evaluate our proximal distance algorithms on the basis of (a) the number of iterations taken, (b) the final value of the penalized objective $f_{\rho}(\bbeta)$, (c) the squared gradient norm $\|\nabla f_{\rho}(\bbeta)\|^{2}$, (d) distance to the constraint set, (e) the number of support vectors, and (f) prediction accuracy on a test set. For each data set, we trained each algorithm 100 times, randomly sampling components of $\bbeta_{0}$ from the standard normal distribution. The sparsity level was fixed such that solutions were at least 50\% sparse. Results are summarized in Figure \ref{fig:1}.
While there is variation across all four evaluation metrics, we do not observe substantial outliers.
The most variable metric is test accuracy, while the remaining metrics are relatively stable.
Interestingly, there is no clear winner between MM and SD. In the remaining experiments we initialize $\bbeta$ based on univariate regression with $y_{i} \sim x_{ij} \beta_{j} + \beta_{p+1}$; that is
\begin{equation}
    \label{eq:initheuristic}
    \beta_{j} = \frac{ \sum_{i=1}^{n} (x_{ij} - \bar{x}_{j}) (y_{i} - \bar{y}) }{ \sum_{i=1}^{n} (x_{ij} - \bar{x}_{j})^{2} },
    \quad
    j = 1,2,\ldots,p;
    \qquad
    \beta_{p+1} = \bar{y}.
\end{equation}

\subsection{Prediction and Model Selection}

Our motivation for incorporating sparsity in the model (\ref{eq:basic-primal-model}) is to build feature selection into SVMs directly. Thus, we are interested in studying both predictive accuracy and estimation capability. Specifically, we claim that both MM and SD are capable of recovering a sparse model $\bbeta$
with reasonable training data $(y_{i}, \bx_{i})$. These advances improve SVM's predictive power and our understanding of the critical features of the data.

To address these issues, we simulate $(\by, \bX, \bbeta)$ as follows:
\begin{itemize}
    \item[(a)] The number of true causal features is fixed at $k_{0} = 50$.
    \item[(b)] Feature vectors $\bx_{i}$ are drawn independently from the multivariate normal distribution $\mathcal{N}(\bzero, \bI)$.
    \item[(c)] Model parameters corresponding to a causal features, $\beta_{j}$, are selected uniformly from the union $[-10,-2] \cup [2, 10]$. The remaining components are set to 0.
    \item[(d)] Labels $y_{i}$ are simulated under the binary decision rule $\sgn(\bx_i^\top\bbeta)$. 
    \item[(e)] The numbers of samples $n$ and features $p$ are varied to cover different extremes of high-dimensional data.
\end{itemize}
In each problem instance, we determine predictive accuracy from the proportion of correctly classified samples in a test set and assess a binary classifier's ability to select informative features using false discovery and omission rates (FDR and FOR). For completeness, the formulas
\begin{eqnarray*}
    FDR &= \frac{(1-SPC) \times (1-q)}{(1-SPC) \times (1-q) + SEN \times q} \\
    FOR &= \frac{(1-SEN) \times q}{(1-SEN) \times q + SPC \times (1-q)}
\end{eqnarray*}
yield the desired quantities, where sensitivity and specificity (SEN and SPC, respectively) are determined by comparing the fitted coefficients $\hat{\bbeta}$ against the true coefficients $\bbeta$.
Here $q = k_{0} / n$, the proportion of causal features, adjusts both FDR and FOR to allow comparisons under different ground truths for sparsity. Note that we do not use warm starts in solving for a specific sparsity level. 

\begin{figure}[htp]
    \centering
    \includegraphics[width=0.5\textwidth]{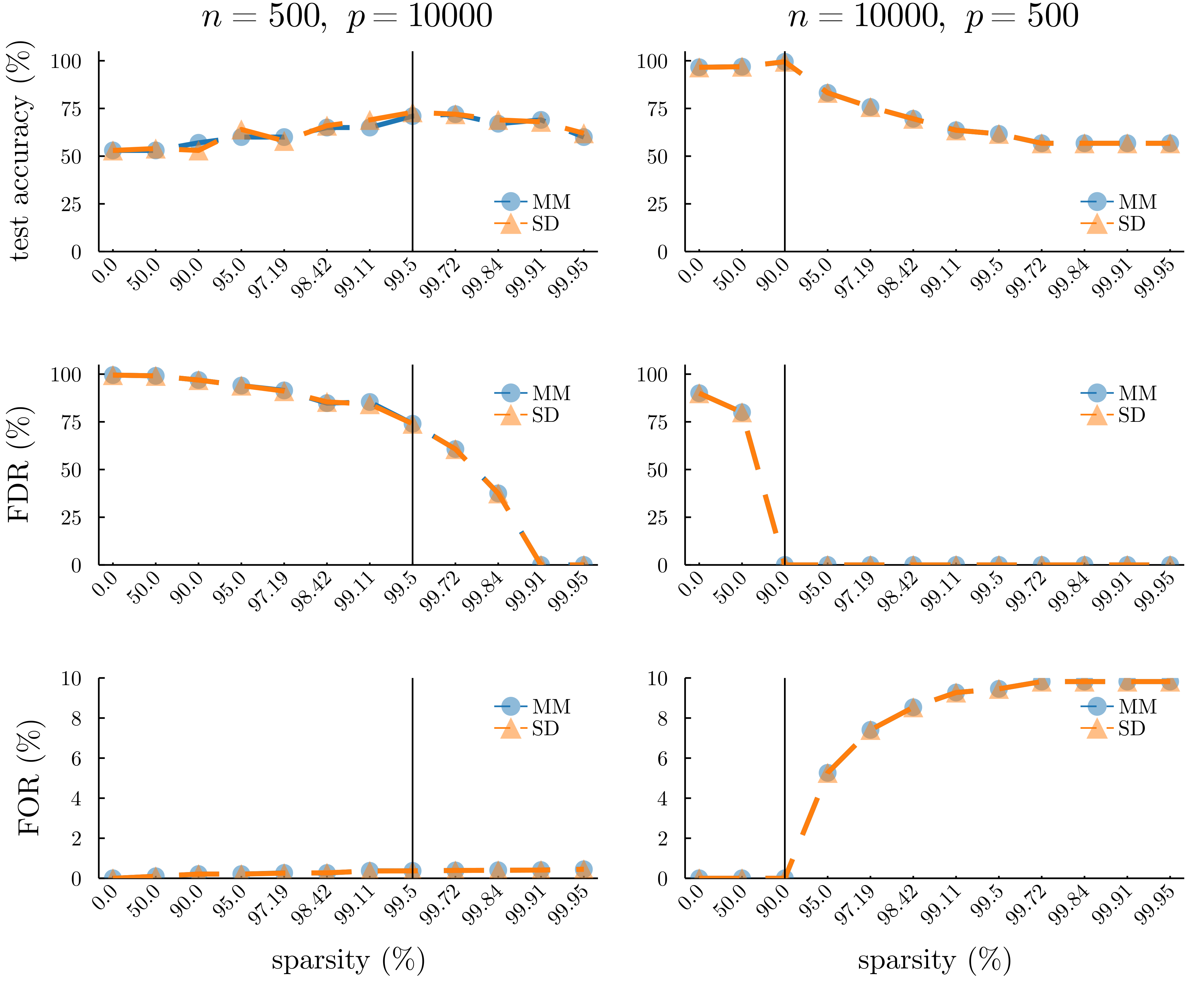}%
    \includegraphics[width=0.5\textwidth]{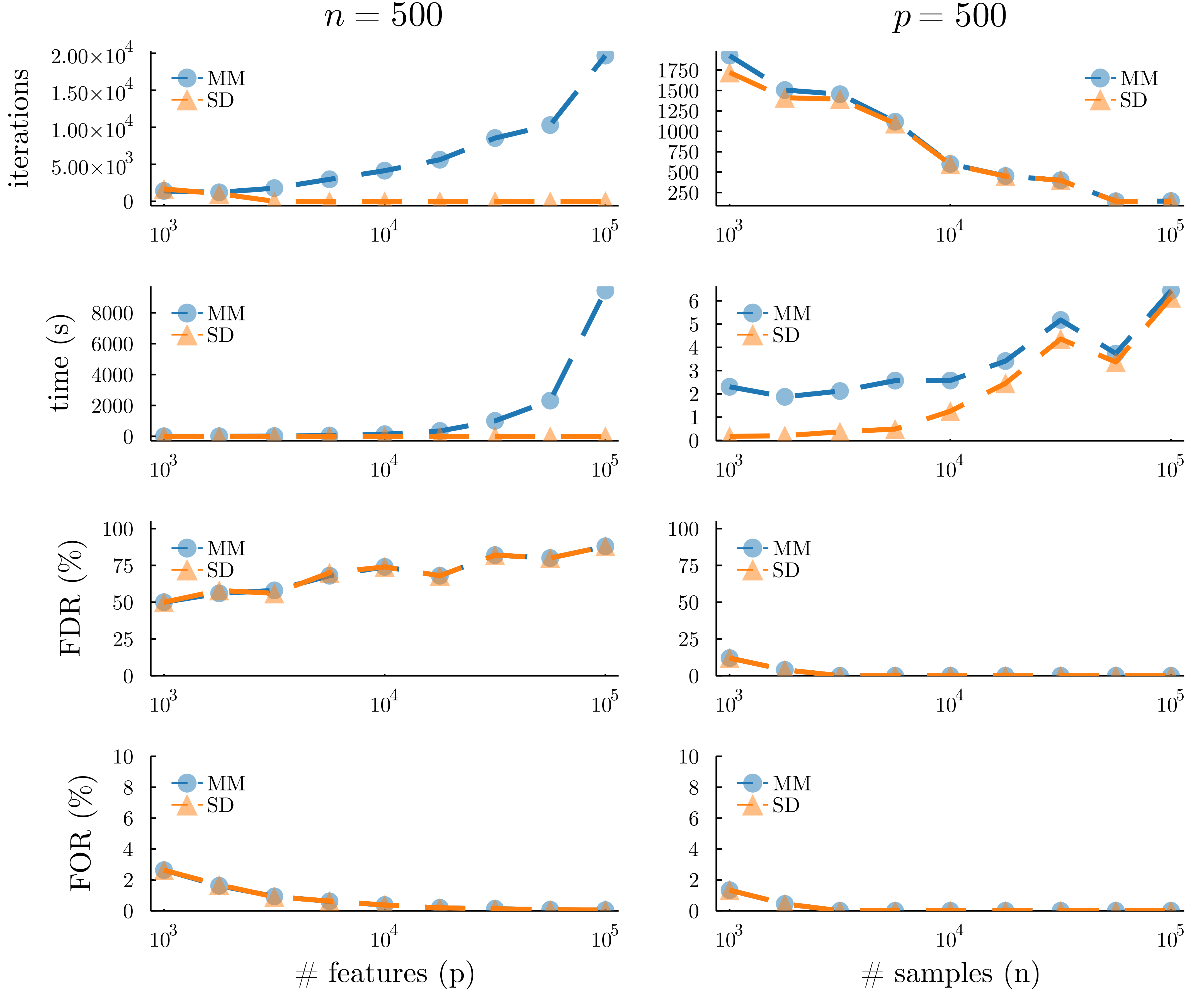}
    \caption{
        Predictive and estimation capabilities of algorithms MM (blue) and SD (orange) across high-dimensional scenarios.
        The black vertical line in the left panels highlights the true sparsity level, $100 \times (1-k_{0} / n)$.
        In the right-hand panels, we report representative results based on the true sparsity level to demonstrate scaling behavior in an ideal regime.
        \label{fig:2}
    }
\end{figure}

Figure \ref{fig:2} summarizes our results.
Both MM and SD have similar performance characteristics as feature or sample dimensions increase. In general, our algorithms minimize FOR and FDR near the true sparsity level, and both error metrics increase if one insists on too simple of a model. FOR is consistently low, demonstrating that the methods are good at eliminating uninformative features, especially as the number of training samples increases.
However, FDR may be inflated in the underdetermined regime $(n=500, p=1000)$ if data are too noisy. We attribute this result to the large number of uncorrelated features $\bx_{i}$ simulated. Another potential issue is that $\bbeta$ is initialized with the heuristic  (\ref{eq:initheuristic}) rather than reusing previous solutions as warm starts. We consider warm starts in the next experiments.

\subsection{Cross-Validation}

\begin{figure}[htp]
    \centering
    \includegraphics[width=0.48\linewidth]{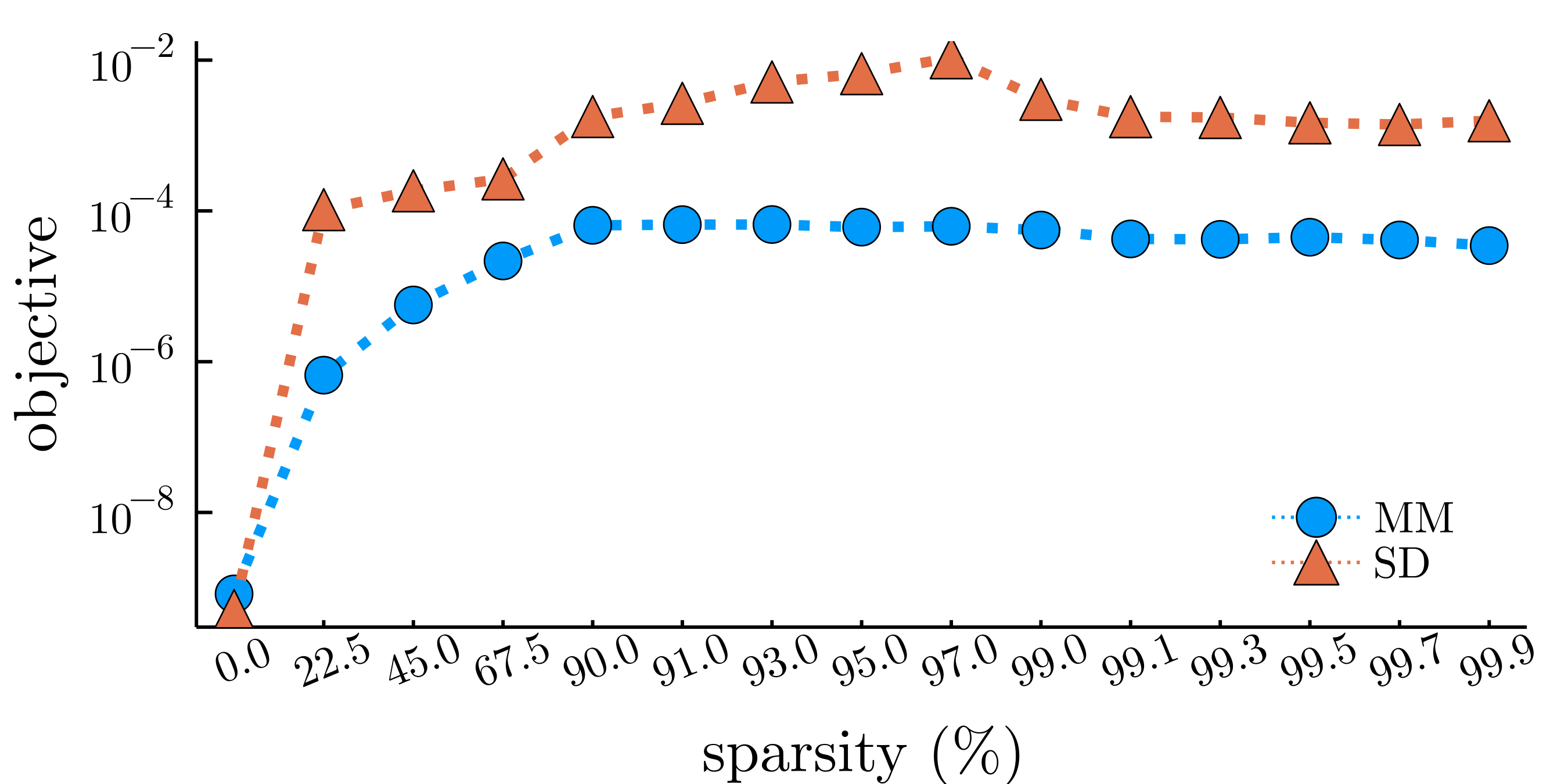}%
    \includegraphics[width=0.48\linewidth]{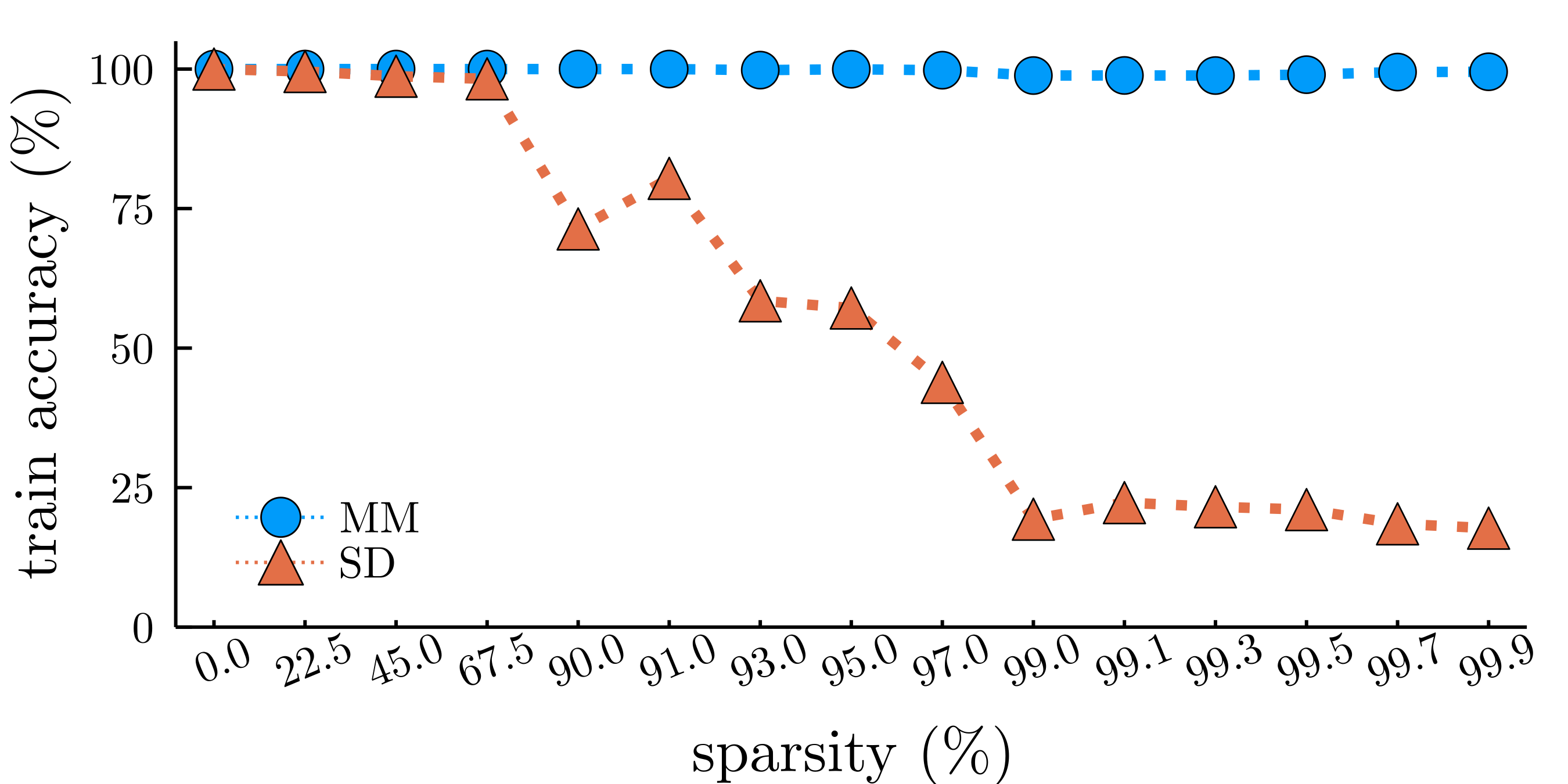}

    \includegraphics[width=0.48\linewidth]{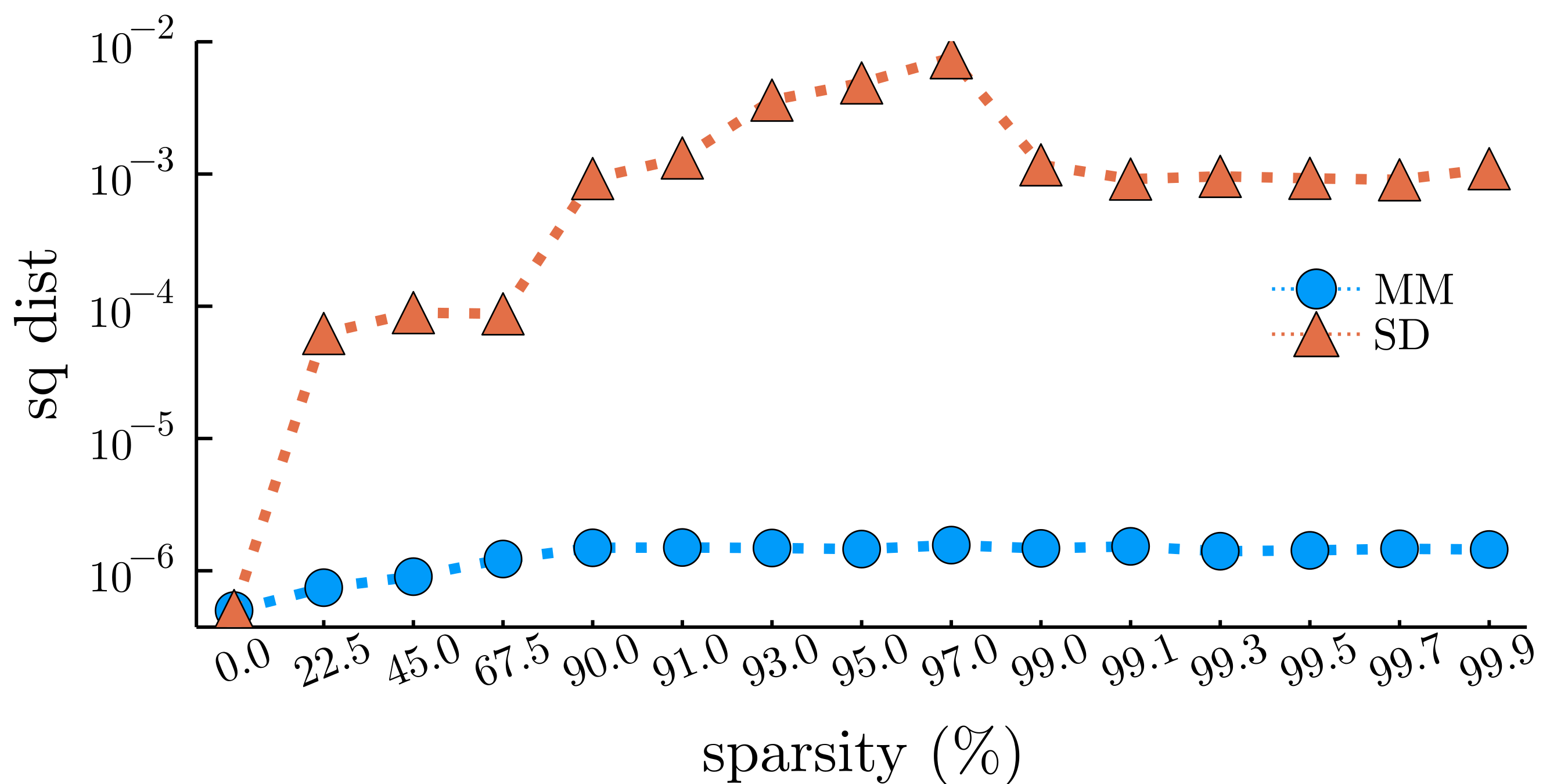}%
    \includegraphics[width=0.48\linewidth]{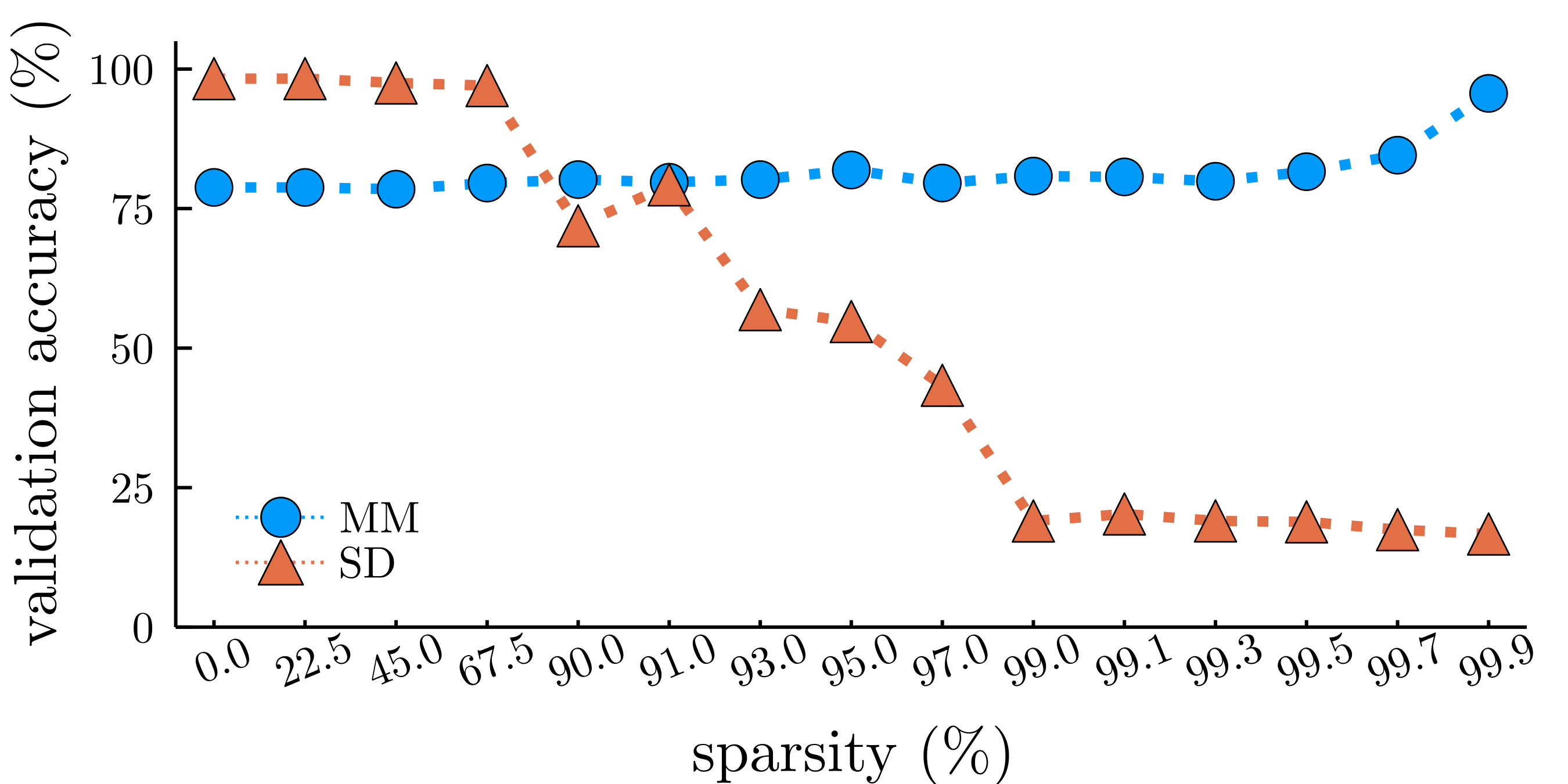}

    \includegraphics[width=0.48\linewidth]{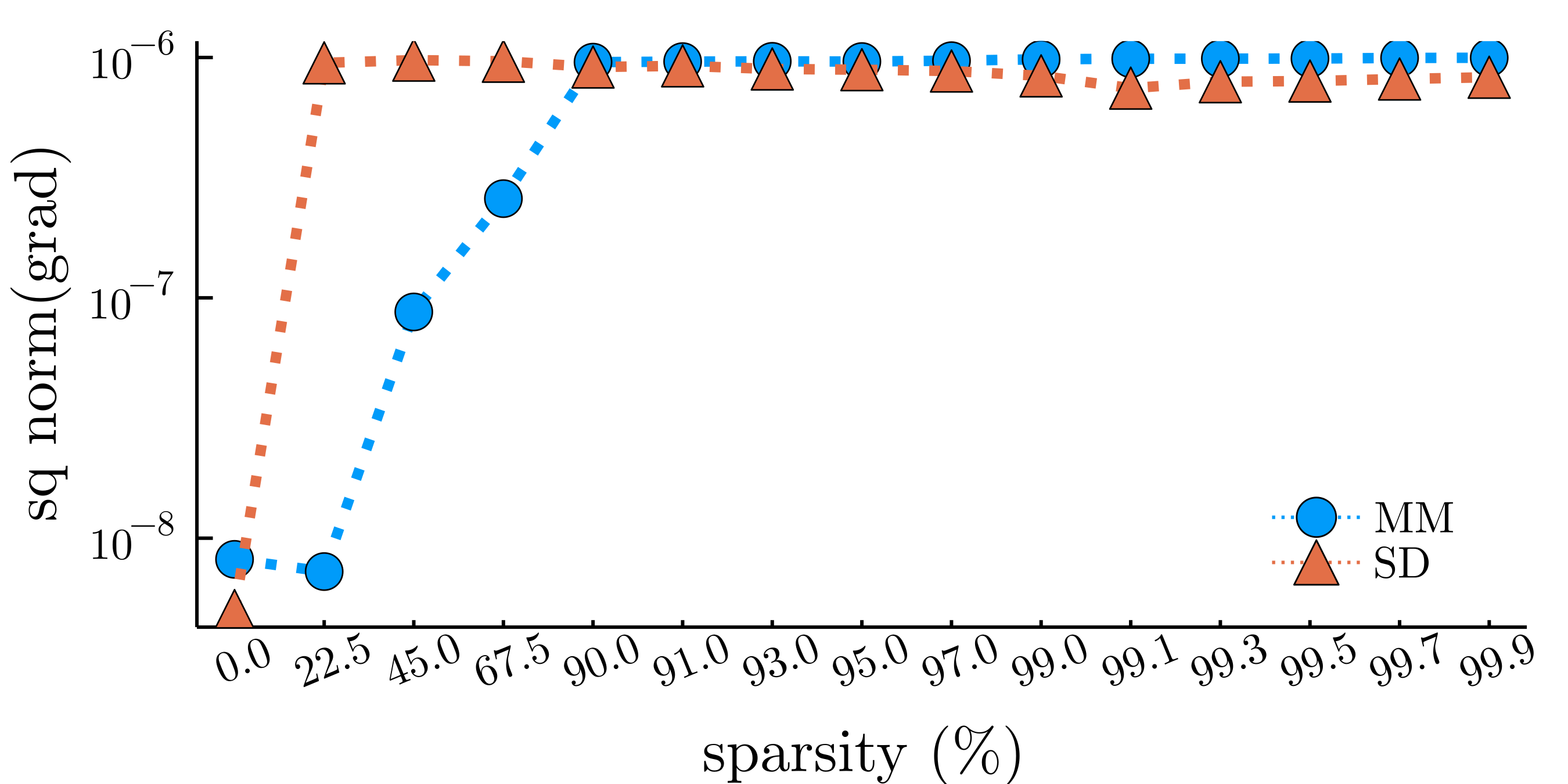}%
    \includegraphics[width=0.48\linewidth]{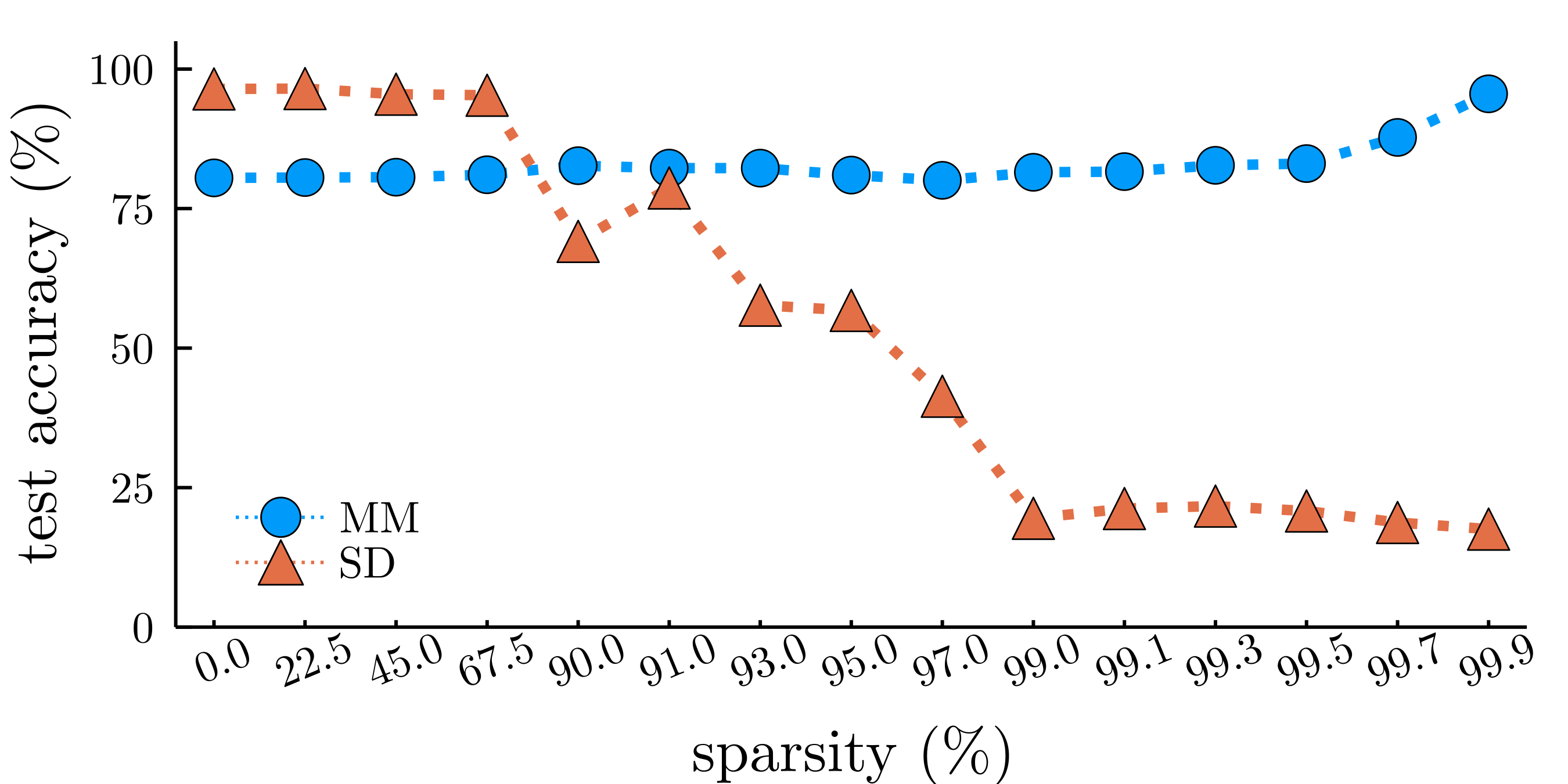}

    \caption{
        Convergence metrics (left) and evaluation metrics (right) for Algorithms MM and SD on the \texttt{TCGA-PANCAN-HiSeq} example.
        \label{fig:3}
    }
\end{figure}

Without prior knowledge of the true number of causal features, one must tune the sparsity level of the solutions. Here we study compatibility of $10$-fold cross-validation of our algorithms with warm starts.
Specifically, in each fold we initialize $\bbeta$ with the heuristic (\ref{eq:initheuristic}) and then solve (\ref{eq:basic-primal-model}) without any sparsity constraints. Then solution sparsity $s = 1-k/n$ is slowly increased from fully dense (0\%) to fully sparse (100\%) by solving the distance penalized version of the basic soft-margin model. Let us briefly clarify our terminology:
\begin{itemize}
    \item[(a)] A training set (Tr) is used to fit model coefficients in each fold.
    \item[(b)] A validation set (V) is used to tune the hyperparameter $k$ (or equivalently, $s$) based on prediction accuracy within each fold.
    \item[(c)] A test set (T) is used to evaluate performance of selected models. This set contains samples never encountered in the training or test sets.
\end{itemize}
We report selected evaluation metrics in the following tables, aggregated as averages over cross-validation folds. Table \ref{tab:2} illustrates performance of MM on the \texttt{synthetic} example in detail, and Table \ref{tab:3} reports the highlights across all 8 examples.
\begin{table}[htp]
    \tiny
    \centering
    \begin{tabular}{crrrrrrrr}
\toprule
$s$ (\%) & Iter. & Time (s) & Objective & Squared Distance & Train (\%) & Valid. (\%) & Test (\%) & SV\\
\midrule
$0$ & $15.9$ & $0.04178$ & $4.559 \cdot 10^{-07}$ & $0$ & $100$ & $73$ & $76.5$ & $26.7$\\
$22.5$ & $49.4$ & $0.06582$ & $1.122 \cdot 10^{-05}$ & $1.207 \cdot 10^{-06}$ & $100$ & $74$ & $76.9$ & $106.5$\\
$45$ & $163.5$ & $0.2091$ & $2.873 \cdot 10^{-05}$ & $1.178 \cdot 10^{-06}$ & $100$ & $74.5$ & $78.4$ & $125.2$\\
$67.5$ & $358.6$ & $0.4365$ & $8.43 \cdot 10^{-05}$ & $1.304 \cdot 10^{-06}$ & $100$ & $80.12$ & $82.5$ & $112.7$\\
$90$ & $839.4$ & $1.024$ & $0.000684$ & $1.52 \cdot 10^{-06}$ & $100$ & $92.5$ & $93.4$ & $65.4$\\
$91$ & $46.9$ & $0.06743$ & $9.166 \cdot 10^{-05}$ & $1.411 \cdot 10^{-05}$ & $100$ & $93.12$ & $93.25$ & $66.2$\\
$93$ & $483.4$ & $0.6224$ & $0.001303$ & $1.501 \cdot 10^{-06}$ & $100$ & $94.88$ & $94.55$ & $62.3$\\
$95$ & $545.2$ & $0.6822$ & $0.002178$ & $1.448 \cdot 10^{-06}$ & $100$ & $96.38$ & $95.75$ & $58.7$\\
$97$ & $670.7$ & $0.8424$ & $0.003874$ & $1.368 \cdot 10^{-06}$ & $99.96$ & $98.12$ & $96.4$ & $57.3$\\
$99$ & $788.8$ & $0.9689$ & $0.007745$ & $1.323 \cdot 10^{-06}$ & $99.73$ & $99.25$ & $98.35$ & $59.5$\\
$99.1$ & $219$ & $0.2715$ & $0.00758$ & $1.421 \cdot 10^{-06}$ & $99.73$ & $99.25$ & $98.35$ & $59.2$\\
$99.3$ & $275.8$ & $0.3419$ & $0.008007$ & $1.423 \cdot 10^{-06}$ & $99.72$ & $99.12$ & $98.35$ & $58.8$\\
$99.5$ & $361.6$ & $0.4409$ & $0.009339$ & $1.364 \cdot 10^{-06}$ & $99.65$ & $99$ & $99$ & $60.4$\\
$\textbf{99.7}$ & $\textbf{471.3}$ & $\textbf{0.5772}$ & $\textbf{0.01005}$ & $\textbf{1.293} \cdot 10^{-06}$ & $\textbf{99.87}$ & $\textbf{99.88}$ & $\textbf{99.5}$ & $\textbf{61.7}$\\
$99.9$ & $1072$ & $1.334$ & $0.2693$ & $1.048 \cdot 10^{-06}$ & $79.61$ & $79.5$ & $82.75$ & $566.7$\\
\bottomrule
\end{tabular}

    \caption{
        Results for 10-fold cross-validation on \texttt{synthetic} data set using MM at various sparsity levels $s$.
        The highlighted row corresponds to results for the true sparsity level.
        \label{tab:2}
    }
\end{table}
Accuracy across training, validation, and test sets peak around the true sparsity level $s=99.7\%$ in the \texttt{synthetic} example. Table \ref{tab:2} shows that MM can quickly explore different sparsity levels and that distance-to-sparsity sets is a compelling form of regularization for SVMs. In Table \ref{tab:3} we observe that the combination of our proximal distance algorithms with cross-validation can select parsimonious, predictive models across a wide range of examples. The MM and SD algorithms share similar performance characteristics, but the latter tends to deliver solutions faster due to its cheap computational cost per iteration.
However, the two methods diverge in fitting quality solutions on higher dimensional datasets, namely the \texttt{optdigits} ($5620 \times 64$), \texttt{letter-recognition} ($20000 \times 16$), and \texttt{TCGA-PANCAN-HiSeq} ($801 \times 10000$).
Specifically, MM and SD may select models with differing sparsity levels under cross-validation.
The difference stems from slow progress in driving the distance penalty downhill under the SD algorithm as reported in Figure \ref{fig:3}.
Indeed, the \texttt{TCGA-PANCAN-HiSeq} example demonstrates that SD may fail to fit training data under our chosen stopping rules which are designed to limit iterations in the face of slow convergence.
Poor adherence to a specified constraint set $S_{k}$ ultimately leads SD astray in model selection.
\begin{table}[htp]
    \tiny
    \centering
    \begin{tabular}{cccrrrrr}
\toprule
Dataset & Alg. & $s$ (\%) & Total Iter. & Total Time (s) & Objective & V (\%) & T (\%)\\
\midrule
\texttt{TCGA-PANCAN-HiSeq} & MM & $99.9$ & $9.181 \cdot 10^{4}$ & $1759$ & $3.467 \cdot 10^{-05}$ & $95.64$ & $95.56$\\
\texttt{} & SD & $0$ & $8.114 \cdot 10^{5}$ & $691.4$ & $0$ & $98.29$ & $96.44$\\
\midrule
\texttt{breast-cancer-wisconsin} & MM & $30$ & $4461$ & $0.01484$ & $0.05329$ & $96.53$ & $99.49$\\
\texttt{} & SD & $30$ & $8124$ & $0.02265$ & $0.05329$ & $96.53$ & $99.42$\\
\midrule
\texttt{iris} & MM & $25$ & $858.1$ & $0.0008536$ & $0.0015$ & $96.67$ & $81.66$\\
\texttt{} & SD & $25$ & $1205$ & $0.0009029$ & $0.001623$ & $96.67$ & $81.33$\\
\midrule
\texttt{letter-recognition} & MM & $0$ & $5.152 \cdot 10^{5}$ & $6.419$ & $0.02848$ & $82.07$ & $81.06$\\
\texttt{} & SD & $12.5$ & $8.788 \cdot 10^{5}$ & $9.552$ & $0.02869$ & $81.96$ & $80.97$\\
\midrule
\texttt{optdigits} & MM & $0$ & $1.026 \cdot 10^{6}$ & $37.23$ & $5.149 \cdot 10^{-08}$ & $95.71$ & $93.6$\\
\texttt{} & SD & $22.5$ & $3.989 \cdot 10^{6}$ & $83.54$ & $3.309 \cdot 10^{-06}$ & $97.15$ & $95.47$\\
\midrule
\texttt{spiral} & MM & $93$ & $4.68 \cdot 10^{4}$ & $54.7$ & $1.955 \cdot 10^{-05}$ & $99.62$ & $100$\\
\texttt{} & SD & $45$ & $4.598 \cdot 10^{4}$ & $2.553$ & $1.655 \cdot 10^{-05}$ & $99.75$ & $100$\\
\midrule
\texttt{splice} & MM & $93$ & $1.979 \cdot 10^{4}$ & $3.896$ & $0.03915$ & $95.21$ & $94.96$\\
\texttt{} & SD & $90$ & $3.17 \cdot 10^{4}$ & $1.762$ & $0.03544$ & $94.94$ & $94.77$\\
\midrule
\texttt{synthetic} & MM & $99.7$ & $6362$ & $7.927$ & $0.01005$ & $99.88$ & $99.5$\\
\texttt{} & SD & $99.7$ & $5204$ & $0.5066$ & $0.0001235$ & $99.75$ & $99.35$\\
\bottomrule
\end{tabular}

    \caption{
        Summary of 10-fold cross-validation results. Column V is the average validation set accuracy used to select a model in cross-validation. The selected model is then evaluated on a test set T.  Total iteration count and total time are based on solution path traversal from $s = 0\%$ to $s = 100\%$.
        \label{tab:3}
    }
\end{table}

\subsection{Comparison to LIBSVM}

We compare our algorithms to algorithms  implemented in the Julia version LIBSVM.jl of LIBSVM \cite{chang2011libsvm} and LIBLINEAR \cite{fan2008liblinear}. Specifically we compare MM and SD to:
\begin{itemize}
    \item[(a)] L2R, an option from LIBLINEAR for the standard $\ell_{2}$ regularized version of criterion (\ref{eq:basic-primal-model}).
    \item[(b)] L1R, an option from LIBLINEAR for the $\ell_{1}$ regularized version of (\ref{eq:basic-primal-model}).
    \item[(c)] SVC, an option from LIBSVM chosen specifically for its nonlinear example \texttt{spiral}. SVC is run using a Gaussian kernel and assumes the standard hinge-loss model.
\end{itemize}
Table \ref{tab:4} reports test set accuracy after 10-fold cross-validation to select the sparsity hyperparameter $(s)$ in our methods and the classifier margin ($C$) in the LIBSVM methods. Both LIBLINEAR and LIBSVM classifiers rely on the OVO paradigm for multiclass assignment. The final column in Table \ref{tab:4} reports the observed sparsity in the final model. If the classification task involves $c>2$ classes, then we average observed sparsity levels of the non-intercept terms in $\bbeta$ over each of the $\binom{c}{2}$ SVMs. Our SVM training algorithms are comparable to existing approaches, albeit slower in the examples \texttt{TCGA-PANCAN-HiSeq} and \texttt{spiral}. To our credit, the \texttt{synthetic} example underscores the superiority of sparsity constraints over shrinkage-based penalties.
\begin{table}[htp]
    \tiny
    \centering
    \begin{tabular}{cccrrrrr}
\toprule
Dataset & Alg. & $s$ (\%) or $C$ & Total Time (s) & Tr (\%) & V (\%) & T (\%) & Sparsity\\
\midrule
\texttt{TCGA-PANCAN-HiSeq} & MM & $99.9$ & $1750$ & $99.52$ & $95.64$ & $95.56$ & $\textbf{99.9}$\\
\texttt{} & SD & $0$ & $766.1$ & $\textbf{100}$ & $98.29$ & $96.44$ & $0$\\
\texttt{} & L2R & $1$ & $14.1$ & $\textbf{100}$ & $\textbf{99.84}$ & $\textbf{100}$ & $0.066$\\
\texttt{} & L1R & $1$ & $\textbf{10.35}$ & $\textbf{100}$ & $\textbf{99.84}$ & $\textbf{100}$ & $99.49$\\
\midrule
\texttt{breast-cancer-wisconsin} & MM & $30$ & $0.01532$ & $\textbf{96.38}$ & $\textbf{96.53}$ & $\textbf{99.49}$ & $\textbf{33.33}$\\
\texttt{} & SD & $30$ & $0.02266$ & $\textbf{96.38}$ & $\textbf{96.53}$ & $99.42$ & $\textbf{33.33}$\\
\texttt{} & L2R & $1$ & $\textbf{0.001334}$ & $85.8$ & $83.68$ & $91.82$ & $0$\\
\texttt{} & L1R & $0.01$ & $0.004505$ & $86.35$ & $84.04$ & $90.8$ & $18.89$\\
\midrule
\texttt{iris} & MM & $25$ & $0.0007963$ & $\textbf{100}$ & $\textbf{96.67}$ & $\textbf{81.66}$ & $\textbf{25}$\\
\texttt{} & SD & $25$ & $0.0009201$ & $\textbf{100}$ & $\textbf{96.67}$ & $81.33$ & $\textbf{25}$\\
\texttt{} & L2R & $1$ & $\textbf{0.0001739}$ & $85.74$ & $81.67$ & $80.66$ & $0$\\
\texttt{} & L1R & $1$ & $0.0002635$ & $86.02$ & $80$ & $80.66$ & $11.67$\\
\midrule
\texttt{letter-recognition} & MM & $0$ & $6.765$ & $\textbf{83.8}$ & $\textbf{82.07}$ & $\textbf{81.06}$ & $0$\\
\texttt{} & SD & $12.5$ & $10.04$ & $83.75$ & $81.96$ & $80.97$ & $\textbf{12.5}$\\
\texttt{} & L2R & $0.775$ & $\textbf{3.318}$ & $68.07$ & $67.49$ & $66.2$ & $0$\\
\texttt{} & L1R & $0.775$ & $41.44$ & $68.04$ & $67.47$ & $66.18$ & $1.106$\\
\midrule
\texttt{optdigits} & MM & $0$ & $36.77$ & $\textbf{100}$ & $95.71$ & $93.6$ & $0$\\
\texttt{} & SD & $22.5$ & $79.26$ & $\textbf{100}$ & $\textbf{97.15}$ & $\textbf{95.48}$ & $21.88$\\
\texttt{} & L2R & $0.005$ & $\textbf{0.9293}$ & $97.86$ & $96.1$ & $94.93$ & $3.281$\\
\texttt{} & L1R & $0.05$ & $4.291$ & $97.74$ & $96.1$ & $95.19$ & $\textbf{37.66}$\\
\midrule
\texttt{spiral} & MM & $90$ & $56.77$ & $99.99$ & $99.62$ & $\textbf{100}$ & $\textbf{90.03}$\\
\texttt{} & SD & $45$ & $2.579$ & $\textbf{100}$ & $99.75$ & $\textbf{100}$ & $45.01$\\
\texttt{} & SVC & $1$ & $\textbf{0.0532}$ & $\textbf{100}$ & $\textbf{100}$ & $\textbf{100}$ & $21.06$\\
\midrule
\texttt{splice} & MM & $93$ & $4.111$ & $96.12$ & $95.21$ & $94.96$ & $\textbf{92.78}$\\
\texttt{} & SD & $90$ & $1.792$ & $96.21$ & $94.94$ & $94.77$ & $90$\\
\texttt{} & L2R & $0.01$ & $\textbf{0.5402}$ & $97.02$ & $95.13$ & $94.55$ & $0$\\
\texttt{} & L1R & $0.03$ & $1.637$ & $\textbf{96.77}$ & $\textbf{96.04}$ & $\textbf{95.09}$ & $61.22$\\
\midrule
\texttt{synthetic} & MM & $99.7$ & $7.836$ & $99.87$ & $\textbf{99.88}$ & $\textbf{99.5}$ & $\textbf{99.6}$\\
\texttt{} & SD & $99.7$ & $0.4528$ & $99.83$ & $99.75$ & $99.35$ & $\textbf{99.6}$\\
\texttt{} & L2R & $0.01$ & $\textbf{0.3411}$ & $\textbf{99.94}$ & $74.25$ & $76.05$ & $0$\\
\texttt{} & L1R & $0.03$ & $0.3606$ & $99.62$ & $98$ & $97.75$ & $90.98$\\
\bottomrule
\end{tabular}

    \caption{
        Comparison of our algorithms (MM and SD) against existing approaches from LIBSVM (L2R, L1R, and SVC). Optimal values for each metric are highlighted with bold text.
        \label{tab:4}
    }
\end{table}

\newpage

\section{Discussion}

We have demonstrated the benefits of conceptually simple proximal distance algorithms for binary and multiclass classificaiton problems on both linear SVMs and nonlinear kernel SVMs. The proximal distance principle  makes it possible to attack parsimony directly through squared distance penalties. This direct approach (a) restores differentiability via quadratic surrogate functions, (b) avoids the shrinkage inherent in lasso-based algorithms, (c) identifies sparser models with good predictive power, and (d) substitutes a discrete interpretable sparsity level for the continuous uninterpretable hyperparameters of competing methods. To our surprise, the more expensive Algorithm MM scales better on high-dimensional data due to its ability to quickly drive solutions close to a desired sparsity set.

While we are pleased with our results, particularly for binary classification tasks, much is left to be desired for multiclass problems. Relying on multiple SVMs to handle multiclass problems introduces $\binom{c}{2}$ subproblems for $c$ true classes. Furthermore, different decision boundaries in the OVO paradigm may be driven by different features, obscuring the universal features that discriminate between classes.

Hence, it is natural to investigate multiclass methods beyond hyperplane separation. Our previous research on multivertex discriminant analysis (MVDA) \cite{lange2008mm} explored a multiclass model that represents classes geometrically as vertices of a regular simplex embedded in Euclidean space rather than binary choices from $\{-1,1\}$. MVDA takes advantage of $\epsilon$-insensitive norms and generalizes to nonlinear classification via the kernel trick \cite{wu2010multicategory}. We plan to revisit MVDA and incorporate sparsity based on the proximal distance principle and possibly Huber hinge errors \cite{burg2016gensvm}. Given the length of the current paper and the many unresolved challenges ahead,
this goal is best left to a future paper. 

\bibliographystyle{abbrv}
\bibliography{references}

\appendix

\section{Derivations}

\subsection{Algorithm MM}

In this section we reduce the MM update
\begin{equation}
    \label{eq:MMupdate}
    \bbeta_{m+1} =
        [a^{2} \bX^{\top} \bX + b^{2} \bI]^{-1}
        (a^{2} \bX^{\top} \bz_{m} + b^{2} \bp_{m})
\end{equation}
to a matrix-vector operations via the Woodbury matrix identity
\[
    (\bA + \bU \bC \bV)^{-1}
    =
    \bA^{-1} - \bA^{-1} \bU (\bC^{-1} + \bV \bA^{-1} \bU)^{-1} \bV \bA^{-1}.   
\]
Taking $\bU \bSigma_{r \times r} \bV^{\top}$ as the thin svd of $\bX$, the matrix inverse in (\ref{eq:MMupdate}) is
\begin{eqnarray*}
    [a^{2} \bV \bSigma^2 \bV^{\top} + b^{2} \bI_{p+1}]^{-1}
    &=&
    b^{-2} \left[
        \bI_{p+1} - b^{-2} \bV (a^{-2} \bSigma^{-2} + b^{-2} \bV^{\top} \bV)^{-1} \bV^{\top}
    \right] \\
    &=&
    b^{-2} \left[
        \bI_{p+1} - b^{-2} \bV (a^{-2} \bSigma^{-2} + b^{-2} \bI_{r})^{-1} \bV^{\top}
    \right] \\
    &=&
    b^{-2} \left[
        \bI_{p+1} - a^{2} \bV (b^{2} \bI_{r} + a^{2} \bSigma^{2})^{-1} \bSigma^2 \bV^{\top} 
    \right].
\end{eqnarray*}
Multiplying $b^{2} \bp_{m}$ by the result above yields
\[
    \bp_{m}
    -
    a^{2} \bV (a^{2} \bSigma^{2} + b^{2} \bI_{r})^{-1} \bSigma^{2} \bV^{\top} \bp_{m}.
\]
Multiplying $a^{2} \bX^{\top} \bz_{m}$ by the same matrix gives
\begin{eqnarray*}
    && a^{2} b^{-2} \left[
        \bV \bSigma \bU^{\top}
        -
        a^{2} \bV (a^{2} \bSigma^{2} + b^{2} \bI_{r})^{-1} \bSigma^{2} \bV^{\top} \bV \bSigma \bU^{\top}
    \right] \bz_{m} \\
    &=&
    a^{2} b^{-2} \bV \left[
        \bI
        -
        a^{2} (a^{2} \bSigma^{2} + b^{2} \bI_{r})^{-1} \bSigma^{2}
    \right] \bSigma \bU^{\top} \bz_{m} \\
    &=&
    a^{2} b^{-2} \bV (a^{2} \bSigma^{2} + b^{2} \bI_{r})^{-1} \left[
        (a^{2} \bSigma^{2} + b^{2} \bI_{r}) - a^{2} \bSigma^{2}
    \right] \bSigma \bU^{\top} \bz_{m} \\
    &=&
    a^{2} \bV (a^{2} \bSigma^{2} + b^{2} \bI_{r})^{-1} \bSigma \bU^{\top} \bz_{m}.
\end{eqnarray*}
Now sum together the two results to yield the desired update
\[
    \bbeta_{m+1}
    =
    \bp_{m}
    +
    a^{2} \bV (a^{2} \bSigma^{2} + b^{2} \bI_{r})^{-1} \left[
        \bSigma \bU^{\top} \bz_{m}
        -
        \bSigma^{2} \bV^{\top} \bp_{m}
    \right].
\]
Computing the diagonal matrices $\bSigma$ and $(a^{2} \bSigma^{2} + b^{2} \bI_{r})^{-1}$ reduces to scalar operations and may be cached.
Thus, the remainder of the calculation reduces to matrix-vector operations (BLAS Level 2).

\subsection{Algorithm SD}

The next iterate in SD is
\[
\bbeta_{m+1} =  \bbeta_m-t_m \nabla f_\rho(\bbeta_m) ,
\]
where $t_m>0$ is an optimally chosen step-length. To find $t_m$, we expand $g_\rho(\bbeta \mid \bbeta_m)$
along the direction $-\nabla f_\rho(\bbeta_m) = -\nabla g_\rho(\bbeta_m \mid \bbeta_m)$. The second-order expansion
\begin{eqnarray*}
g_\rho[\bbeta_m - t_m \nabla f_\rho(\bbeta_m) \mid \bbeta_m]
& = & g_\rho(\bbeta_m) - t_m \nabla f_\rho(\bbeta_m)^\top \nabla f_\rho(\bbeta_m) \\
&  & - \frac{t_m^2}{2}\nabla f_\rho(\bbeta_m)^\top d^2g_\rho(\bbeta_m \mid \bbeta_m)
\nabla f_\rho(\bbeta_m) \\
& = & g_\rho(\bbeta_m) - t_m \|\nabla f_\rho(\bbeta_m)\|^2  \\
&  & + \frac{t_m^2}{2}[a^2\|\bX \nabla f_\rho(\bbeta_m)\|^2+ b^{2} \|\nabla f_\rho(\bbeta_m)\|^2]
\end{eqnarray*}
is exact because the surrogate is quadratic.
Elementary calculus shows that 
\begin{eqnarray*}
t_{m} & = & \frac{\|\nabla f_\rho(\bbeta_m)\|^{2} }{
        a^{2} \|\bX \nabla f_\rho(\bbeta_m)\|^{2} + b^{2} \|\nabla f_\rho(\bbeta_m)\|^2}.
\end{eqnarray*}
maximizes the surrogate and therefore drives the objective $f_\rho(\bbeta)$ downhill.  At a stationary point of $f_\rho(\bbeta)$, both the MM and SD algorithms are also stationary.

\newpage

\section{Simulated Data Sets}

\subsection{\texttt{Synthetic}}

\begin{figure}[htp]
    \centering
    \includegraphics[width=\textwidth]{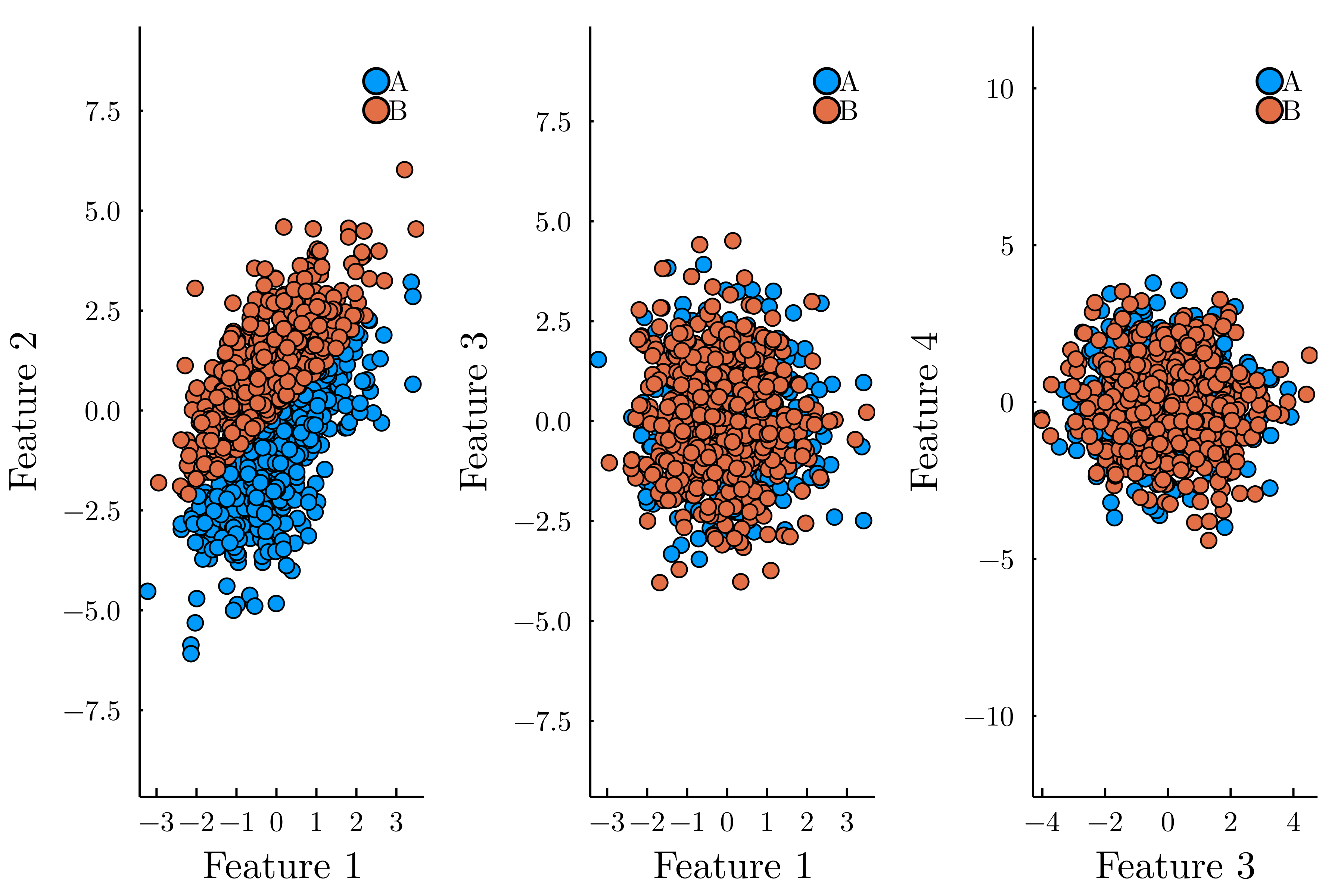}
    \caption{
        Views of the \texttt{synthetic} data set using two informative features (left), an informative and noninformative feature (middle), and two uninformative features (right).
        \label{fig:S1}
    }
\end{figure}

Features $\bx_{i} \in \Real^{p}$ are drawn from a multivariate distribution centered at $\bzero$ and variance-covariance matrix $\bSigma$.
Components of $\bSigma$ are chosen as follows
\begin{eqnarray*}
    \Sigma_{11} &=& 1, \\
    \Sigma_{22} &=& 3, \\
    \Sigma_{12} &=& 0.9, \\
    \Sigma_{jj} &=& 2, \quad \text{for } j=1,2,\ldots p. \\
    \Sigma_{ij} &=& 10^{-3} \mathcal{N}(0,1) \quad \text{otherwise}.
\end{eqnarray*}
The true model $\bbeta$ is given by
\[
    \beta_{1} = 10,
    \quad
    \beta_{2} = -10,
    \quad
    \beta_{j} = 0
    \quad
    \text{otherwise}.
\]
Figure \ref{fig:S1} visualizes our example from multiple perspectives.

\newpage

\subsection{\texttt{Spiral}}
\begin{figure}[htp]
    \centering
    \includegraphics[width=\textwidth]{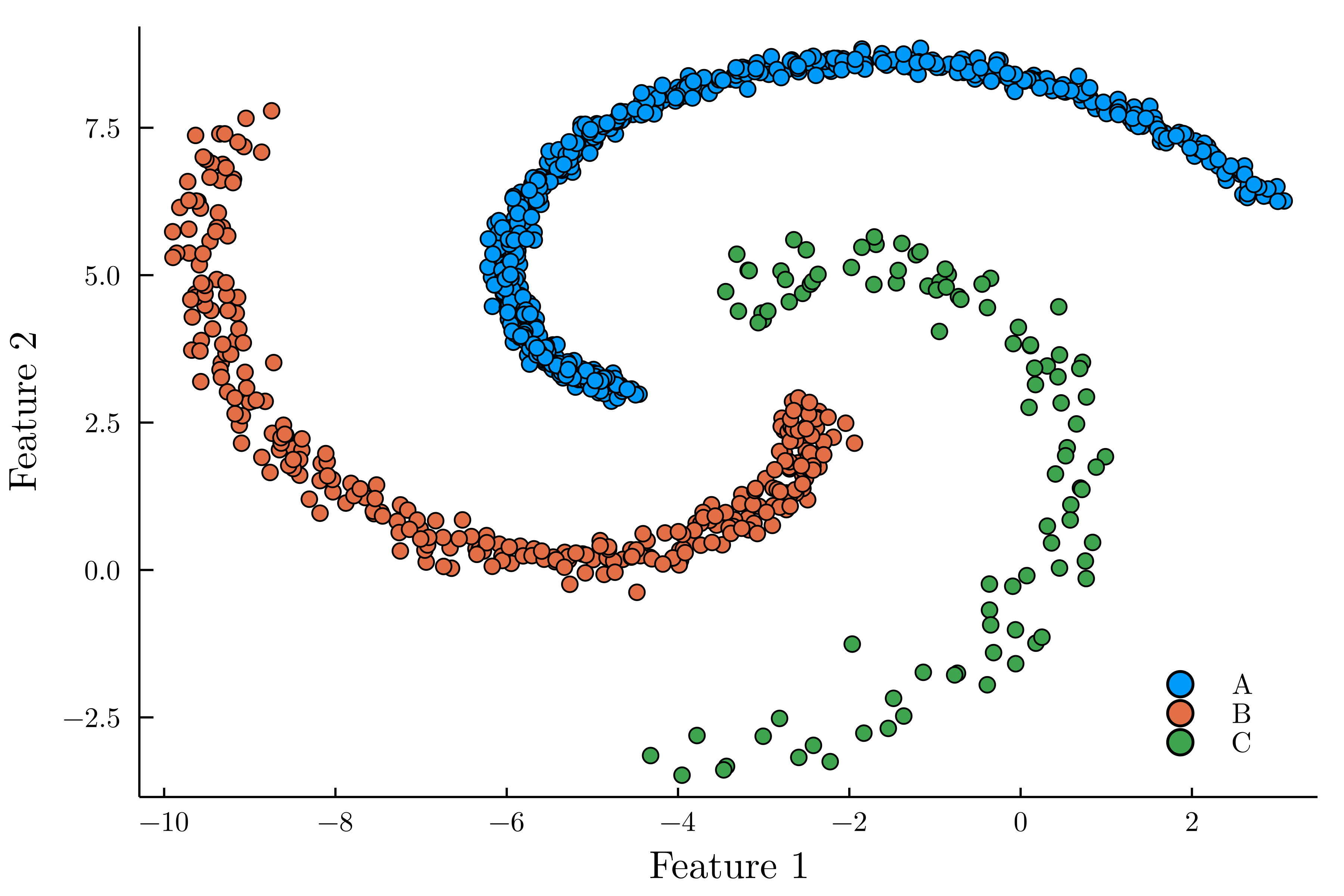}
    \caption{
        Spirals representing three distinct classes.
        \label{fig:S2}
    }
\end{figure}

Starting with $x_{0} = -3.5$ and $y_{0} = 3.5$, coordinates for each spiral sample $i$ are given by
\begin{eqnarray*}
    x_{i1} &\equiv& x_{i} = x_{0} + r_{ic} \cos \theta_{ic}
        + \mathcal{N}(0,\sigma_{c}^{2}), \\
    x_{i2} &\equiv& y_{i} = y_{0} + r_{ic} \sin \theta_{ic}
        + \mathcal{N}(0,\sigma_{c}^{2}),
\end{eqnarray*}
where $c$ indicates class membership.
The sequences of polar coordinates $(r_{ic}, \theta_{ic})$ for each class $c \in \{A,B,C\}$ are defined by the recipes
\begin{align*}
    r_{iA} &= 7 \left(1 - \frac{k_{iA}}{n_{A} + n_{A} / 5} \right),
    &\theta_{iA} &= \frac{\pi}{8} + \frac{k_{iA} \pi}{n_{A}},
    &n_{A} &= 600; \\
    r_{iB} &= 7 \left(1 - \frac{k_{iB}}{n_{B} + n_{B} / 5} \right),
    &\theta_{iB} &= \frac{\pi}{8} + \frac{k_{iB} \pi}{n_{B}} + \frac{2\pi}{3},
    &n_{B} &= 300; \text{ and } \\
    r_{iC} &= 7 \left(1 - \frac{k_{iC}}{n_{C} + n_{C} / 5} \right),
    &\theta_{iC} &= \frac{\pi}{8} + \frac{k_{iC} \pi}{n_{C}} + \frac{4\pi}{3},
    &n_{C} &= 100.
\end{align*}
Here $1 \le k_{ic} \le n_{ic}$ is merely an index within a particular class.
We set $\sigma_{A} = 0.1$, $\sigma_{B} = 0.2$, and $\sigma_{C} = 0.3$.
Figure \ref{fig:S2} illustrates the dataset used in our numerical experiments.

\end{document}